\documentclass[sigconf, screen]{acmart}

\IfFileExists{framed.sty}{\usepackage{framed}}{}
\IfFileExists{xcolor.sty}{\usepackage{xcolor}}{}
\IfFileExists{graphicx.sty}{\usepackage{graphicx}}{}
\definecolor{shadecolor}{RGB}{240,240,240}

\usepackage{url}
\usepackage{colortbl}
\usepackage{multirow} 

\acmSubmissionID{3192}

\AtBeginDocument{%
  \providecommand\BibTeX{{%
    \normalfont B\kern-0.5em{\scshape i\kern-0.25em b}\kern-0.8em\TeX}}}

\copyrightyear{2025}
\acmYear{2025}
\setcopyright{cc}
\setcctype{by}
\acmConference[CHI '25]{CHI Conference on Human Factors in Computing Systems}{April 26-May 1, 2025}{Yokohama, Japan}
\acmBooktitle{CHI Conference on Human Factors in Computing Systems (CHI '25), April 26-May 1, 2025, Yokohama, Japan}\acmDOI{10.1145/3706598.3713572}
\acmISBN{979-8-4007-1394-1/25/04}

\begin{document}

\title[Wearable AR in Everyday Contexts]{Wearable AR in Everyday Contexts: Insights from a Digital Ethnography of YouTube Videos}

\author{Tram Thi Minh Tran}
\email{tram.tran@sydney.edu.au}
\orcid{0000-0002-4958-2465}
\affiliation{Design Lab, Sydney School of Architecture, Design and Planning,
  \institution{The University of Sydney}
  \city{Sydney}
  \state{NSW}
  \country{Australia}
}

\author{Shane Brown}
\email{shane@contxtu.al}
\orcid{0009-0004-3147-7220}
\affiliation{
  \institution{Contxtual}
  \city{Sydney}
  \state{NSW}
  \country{Australia}
}

\author{Oliver Weidlich}
\email{oliver@contxtu.al}
\orcid{0009-0006-5907-2908}
\affiliation{
  \institution{Contxtual}
  \city{Sydney}
  \state{NSW}
  \country{Australia}
}

\author{Soojeong Yoo}
\email{soojeong.yoo@sydney.edu.au}
\orcid{0000-0003-3681-6784}
\affiliation{Design Lab, Sydney School of Architecture, Design and Planning,
  \institution{The University of Sydney}
  \city{Sydney}
  \state{NSW}
  \country{Australia}
}

\author{Callum Parker}
\email{callum.parker@sydney.edu.au}
\orcid{0000-0002-2173-9213}
\affiliation{Design Lab, Sydney School of Architecture, Design and Planning,
  \institution{The University of Sydney}
  \city{Sydney}
  \state{NSW}
  \country{Australia}
}

\renewcommand{\shortauthors}{Tran et al.}

\begin{abstract}

With growing investment in consumer augmented reality (AR) headsets and glasses, wearable AR is moving from niche applications to everyday use. However, current research primarily examines AR in controlled settings, offering limited insights into its use in real-world daily life. To address this gap, we adopt a digital ethnographic approach, analysing 27 hours of 112 YouTube videos featuring early adopters. These videos capture usage ranging from continuous periods of hours to intermittent use over weeks and months. Our analysis shows that currently, wearable AR is primarily used for media consumption and gaming. While productivity is a desired use case, frequent use is constrained by current hardware limitations and the nascent application ecosystem. Users seek continuity in their digital experience, desiring functionalities similar to those on smartphones, tablets, or computers. We propose implications for everyday AR development that promote adoption while ensuring safe, ethical, and socially-aware integration into daily life.

\end{abstract}



\maketitle

\section{Introduction}

The recent launch of the Apple Vision Pro\footnote{February 2, 2024, in the US and worldwide after June 2024.}, along with the introduction of more consumer-oriented smart glasses by major tech companies, marks a pivotal moment in the augmented reality (AR) landscape. These advancements highlight the transition of wearable AR devices from their professional and niche domains (such as industrial~\cite{burova2020utilizing, gattullo2020and, cao2020exploratory, henderson2010exploring} and medical~\cite{gasques2021artemis, eom2022neurolens, baashar2023towards}), into the consumer market. Unlike smartphone-based AR, which requires users to hold and position a device to view digital content, wearable AR integrates digital information directly into the user’s field of view, enabling a more continuous and seamless experience~\cite{starner1997augmented, azuma2019road}.


The growing availability of wearable AR devices suggests a shift towards more pervasive or ubiquitous AR~\cite{bowman2021keynote, grubert2016towards}, where AR technology could become as common as smartphones today. Research on wearable AR has largely focused on advancing display and tracking technologies~\cite{azuma2019road, billinghurst2021grand, tran2023wearable}, interaction paradigms~\cite{lu2020glanceable, lu2021evaluating}, and addressing potential challenges in privacy~\cite{wolf2018we, abraham2024you}, social and ethical issues~\cite{krauss2024makes, mhaidli2021identifying, bonner2023filters}. In terms of practical applications, it is crucial to recognise the dichotomy in AR usage: professional and enterprise AR applications are widely embraced for specific purposes, whereas many consumer AR applications are often seen as ephemeral novelties~\cite{azuma2019road}. 

To address this, \citet{azuma2019road} suggests that compelling AR use cases involve meaningful connections to the real world. \citet{mathis2024everyday} builds on this by examining how AR in conjunction with human intervention, may help users navigate daily challenges. Similarly, \citet{glassner2003everyday} explores potential everyday AR applications, such as locating friends at large events or checking the contents of a fridge without opening it. Research has also looked into virtual monitors as customisable, space-saving alternatives to traditional screens, offering adaptable workspaces~\cite{pavanatto2021we}.

Despite these efforts, a gap remains in understanding how AR integrates into real-world, everyday contexts. With an increasing number of AR headsets and smart glasses entering the market—ranging from affordable options around \$300 to premium models exceeding \$3,500—consumers are documenting their experiences and sharing them on social media platforms. Many share reflections on long-term use, such as in videos titled \textit{`Ray-Ban Meta Smart Glasses After 30 Days,'} or experiments with continuous use, such as \textit{`I Survived 24 Hours In Apple Vision Pro.'} These user-generated videos provide valuable insights into the practical integration of AR into daily routines.

In this study, we employ a digital ethnographic approach to analyse 112 YouTube videos, adapting traditional ethnographic methods to digital environments~\cite{masten2003digital}. By examining these user experiences, we aim to answer the following research questions: \textbf{How are wearable AR devices being used in everyday contexts, and what are the most compelling experiences as perceived by users?}

This paper provides the following contributions:

\begin{itemize}
    \item \textbf{Real-world insights}: Examines everyday use of wearable AR by early adopters and technology reviewers through YouTube content.

    \item \textbf{Dominant and compelling use cases}: Highlights media consumption and gaming as primary use cases, with productivity emerging as a desired but less frequent use case.

    \item \textbf{Practical challenges}: Identifies barriers to prolonged use, including hardware limitations and the lack of robust native AR applications.
    
    \item \textbf{Recommendations}: Highlights implications for advancing everyday AR at both the application and device levels.
    
\end{itemize}

\section{Related Work}

\sloppy This section situates the study within the context of Human-Computer Interaction (HCI) research on everyday wearable AR and the use of digital ethnography as a method to explore how these technologies integrate into daily life.

\subsection{Everyday AR Research}

The field of HCI has long been concerned with understanding and improving human interaction with technologies~\cite{card2018psychology}. Early research focused on graphical user interfaces (GUIs) and tangible interactions, eventually expanding to immersive technologies~\cite{bowman20043d} such as XR (Extended Reality), which encompasses Virtual Reality (VR), Augmented Reality (AR), and Mixed Reality (MR)~\cite{milgram1994taxonomy}.

AR\footnote{AR typically overlays digital elements onto the real world, while MR allows deeper integration and interaction between digital and physical environments~\cite{milgram1994taxonomy}. In this paper, the term `AR' will be used as an umbrella term encompassing both technologies for simplicity, except when discussing specific devices where the distinction is crucial.} has been widely studied in professional contexts, such as industrial maintenance~\cite{burova2020utilizing, gattullo2020and, cao2020exploratory, henderson2010exploring} and healthcare~\cite{gasques2021artemis, eom2022neurolens, baashar2023towards}, where its utility has been demonstrated. Shifting focus to more consumer-oriented use cases, \citet{bowman2021keynote} introduced the concept of \textit{Everyday AR}. This vision imagines a future where virtual displays are ubiquitous, providing constant access to information and applications. Bowman contrasted everyday AR with specialised, niche experiences, arguing that true everyday AR would depend on all-day, always-on AR glasses. This idea aligns with the concept of \textit{Pervasive AR} proposed by \citet{grubert2016towards},
a system that continuously adapts to the user’s changing context, making continuous use feasible. Progress toward this vision of everyday AR has been driven by research directed at several key areas.

Research on wearable AR has focused on overcoming technical challenges related to display technologies, tracking, rendering, and interaction techniques~\cite{azuma2019road, billinghurst2021grand, tran2023wearable}. Efforts include improving visual coherence~\cite{bang2021lenslet, rathinavel2019varifocal, hamasaki2019varifocal}, enhancing large-scale tracking with semantic understanding~\cite{runz2018maskfusion, zhang2019hierarchical}, multimodal rendering~\cite{mandl2021neural, lopes2018adding}, and refining interaction methods (e.g., freehand gestures, speech commands, hardware-based inputs)~\cite{pei2022hand, schmitz2022squeezy, hirzle2019design}. These developments are critical for integrating AR into everyday use seamlessly.

Another major focus has been on designing intuitive and unobtrusive user interfaces for AR. Lu et al.~\cite{lu2020glanceable, lu2021evaluating} introduced Glanceable AR interfaces, enabling users to quickly gather information with brief glances at their periphery. Building on this, \citet{lu2022exploring} explored interface transition mechanisms while \citet{plabst2022push} examined how spatial notification placement impacts user performance. In another important development, \citet{xu2023xair} proposed a design framework incorporating explainable AI to make AR systems’ behaviour more interpretable, reducing user confusion and surprise from unexpected outcomes.

As AR technologies become more accessible, ethical and social concerns grow increasingly important. Privacy issues arise from AR devices collecting and displaying personal information, potentially exposing sensitive activities~\cite{regenbrecht2024see, wolf2018we}. Social acceptability includes concerns about discomfort caused by devices capturing information without bystanders’ knowledge~\cite{denning2014situ, o2023privacy}. The risk of malicious use—such as targeted attacks or manipulative visual overlays~\cite{eghtebas2021advantage, eghtebas2023co}—underscores the need for ethical safeguards to prevent harm. Lastly, AR’s societal impact could also widen divides by providing users with privileged access to information~\cite{regenbrecht2024see}.

While technical challenges, interfaces, and ethical considerations drive the development of AR, understanding how these research efforts translate into everyday applications is equally important.

\subsection{Everyday AR Applications}

Everyday AR applications are designed for regular, repeated use in daily routines. Drawing on the successes of mobile AR applications like Pokémon Go and social camera tools, \citet{azuma2019road} observed that while these examples achieved sustained engagement, most consumer AR applications failed to maintain user interest beyond initial curiosity.

To address this challenge, Azuma argued that compelling everyday AR applications must seamlessly blend real and virtual elements in meaningful ways~\cite{azuma2019road}. By `meaningful,' he referred to experiences where virtual content was deeply connected to the real world, making it feel relevant and integral. He proposed strategies such as reinforcing existing interactions, reskinning the real world with virtual overlays, and helping users remember through persistent digital content~\cite{azuma2015location}. Building on this, researchers like \citet{mathis2024everyday}, who surveyed 60 participants, highlighted promising use cases for assistive interfaces that help users recall information, disconnect from reality, and enhance communication through visual augmentations. \citet{glassner2003everyday} similarly envisioned wearable AR assisting with small, casual tasks, making life easier through graphics-based, text-free applications. In terms of productivity, \citet{pavanatto2021we, pavanatto2024multiple} demonstrated how wearable AR could overcome physical space limitations by using virtual monitors to create flexible, adaptable workspaces. In industry, Meta advanced everyday AR applications with its Aria Everyday Activities Dataset~\cite{lv2024aria}, which included 143 daily activity sequences recorded by multiple wearers across five geographically diverse indoor locations. The dataset was aimed at developing \textit{`truly personalised and contextualised AI assistants that can act as an extension to the wearer’s mind.'}

These studies and industry efforts collectively envision how AR will increasingly play a role in everyday life. Since 2022, affordable consumer AR devices have made this technology accessible to a broader audience beyond specialists and study participants. Meta's Ray-Ban smart glasses exemplify this trend, shipping over 700,000 pairs since their October 2023 release~\cite{wsj2023}. This shift opens up new opportunities to study how wearable AR is being used in daily life, especially during this early phase of adoption. Our study leverages this unique moment to explore real-world integration of wearable AR.

\subsection{Digital Ethnography} 

Digital or online ethnography, introduced by Masten and Plowman in 2003~\cite{masten2003digital}, extends traditional ethnographic methods to digital platforms, enabling observations beyond geographical and temporal boundaries. It involves collecting participants' experiences shared through text, images, and audio to interpret their relevance in daily life. This method has been widely adopted in HCI research, focusing on exploring interactions with emerging technologies and real-life contexts. Examples include the iPhone 3G~\cite{blythe2009critical}, insertable devices~\cite{komkaite2019underneath}, and interactions with robots in public spaces~\cite{nielsen2023using, yu2024understanding}. Platforms like YouTube and TikTok are common data sources, offering insights into user interactions through video content and/or community comments. Similarly, wearable AR is currently in the early adopter phase of the technology adoption lifecycle~\cite{rogers2014diffusion}, where individuals are enthusiastic about exploring new technologies and are more likely to share their experiences online. These early adopters play a critical role in shaping perceptions and influencing the early majority~\cite{rogers2014diffusion}. This presents an opportunity to capture real-world usage patterns that are challenging to observe in controlled settings.

While digital ethnography provides valuable insights, it has limitations. The lack of demographic information about content creators~\cite{nielsen2023using} and the possibility that they may not represent the broader user population~\cite{blythe2009critical} limit the transferability of findings. Content can include satire~\cite{blythe2009critical}, which is difficult to analyse due to its ambiguous intent. Additionally, some content may be performative, staged, or selectively shared by creators~\cite{paay2015connecting}, potentially misrepresenting natural behaviours. To address these issues, we critically evaluate data authenticity and interpret findings within broader patterns observed across multiple videos.

\textit{Ethical Considerations Statement}: Respecting the privacy of individuals sharing content on social media is essential in digital ethnography~\cite{nielsen2023using}. In compliance with European Union’s General Data Protection Regulation (GDPR)~\cite{GDPR2018}, personal data such as usernames were not published. While fair use policies (e.g., YouTube~\cite{FairUse2021}) permit the use of publicly uploaded material for research, informed consent was obtained for quoting content or using identifiable information like screenshots. This study received ethical approval from the University of Sydney Human Research Ethics Committee (HREC), protocol 2024/HE000889.

\subsection{Summary}

While HCI research on everyday AR has advanced in areas such as technical development, interaction design, ethics, and potential applications, much of this work is based on insights from controlled settings. These studies often capture only specific aspects of AR use, offering a limited understanding of how AR technology, as a whole, integrates into real-world consumer life over time. This study addresses this gap by using a digital ethnographic approach to leverage (1) the recent surge in consumer AR headsets and glasses and (2) the experiences documented by early adopters. By observing real-world interactions, this research provides HCI with empirical insights into sustained usage patterns and practical challenges, advancing our understanding of how wearable AR can be meaningfully integrated into daily life.

\section{Methodology}

We selected YouTube for this digital ethnography as it is the largest video-sharing platform~\cite{shepherd2024youtube}, where users often document extended technology experiences. Its ability to host long videos makes it ideal for capturing sustained interactions with wearable AR, unlike platforms like TikTok and Instagram, which focus on short, curated clips. YouTube’s established use in HCI research~\cite{blythe2009critical, komkaite2019underneath, paay2015connecting} further supports its suitability for gathering rich, contextual data on user interactions.

An exploratory search was conducted to understand YouTube’s search functions, the types of videos about wearable AR, and their titles. A notable trend was the focus on usage timelines, such as 7 days or 3 months, reflecting concerns about the longevity and utility of these devices after the initial excitement fades. This preliminary investigation informed the development of video selection criteria and the search strategy.

\subsection{Video Selection Criteria}

\subsubsection{Inclusion Criteria}

Videos were included if they:

\begin{enumerate}
    \item Explicitly mentioned using a wearable AR device continuously for several hours or intermittently over days/months, demonstrating real-world scenarios.
    \item Featured consumer-oriented headsets or smart glasses released between June 2022 and June 2024. The release date criterion is justified by the technological advancements and market readiness observed in this period, when major companies like Apple and Meta announced or released consumer-ready AR glasses and headsets. \autoref{appendix:devices} presents the list of devices meeting these criteria.
\end{enumerate}

\subsubsection{Exclusion Criteria}

Videos were excluded based on:

\begin{itemize}

\item \textit{Format}: Excluded YouTube Shorts (<60 seconds) to focus on substantial content.

\item \textit{Content}: Excluded first-time experiences, general reviews, podcasts, or tutorials that did not reflect genuine, long-term use.

\item \textit{Language}: Excluded non-English videos due to the authors' shared proficiency in English.

\item \textit{Source}: Excluded manufacturer-produced or sponsored videos to minimise bias, though affiliate links in video descriptions were accepted as they are common and unlikely to compromise authenticity.

\end{itemize}

\subsection{Search and Screening Strategy}

\subsubsection{Keywords} 
We defined keywords and phrases that: (1) indicate the duration of usage, such as \textit{hrs}, \textit{hour(s)}, \textit{day(s)}, \textit{week(s)}, \textit{month(s)}, \textit{day in the life}, \textit{realistic/real day}, and \textit{long-term}, and (2) indicate contexts that necessitate longer hours of use, such as \textit{travel}, \textit{work}, and \textit{productive/productivity}. These terms were combined with the device names to form search queries, for example, `Apple Vision Pro hours' and `Rokid Max days'.

\subsubsection{Initial Screening} 

Search results were sorted by YouTube’s default relevance algorithm, which ranks videos using factors like keyword matching and semantic understanding. For each keyword combination, videos were screened based on their titles, thumbnails, descriptions, and chapter highlights to determine eligibility against the predefined criteria. Eligible videos were recorded along with their links and titles.

When eligibility could not be determined solely from metadata, the video was retained for further content screening. To maintain efficiency, we halted searches after 30 consecutive results without finding new eligible videos, a threshold informed by similar studies analysing online video content~\cite{yu2024understanding}. This stage identified \textbf{305} videos for the next phase.

\subsubsection{Content Screening}

Content screening involved a two-step process. First, transcripts were extracted using a Python script with the YouTube Transcript API (see \autoref{appendix:scripts}) and reviewed to assess whether the content met the predefined criteria. The transcripts featured user narrations describing interactions with wearable AR devices, providing insights into subjective experiences and internal states—contrasting with traditional video content, which often consists solely of recorded footage. Second, videos were skimmed to verify transcript findings and capture additional visual or contextual details that were not apparent from the text alone.

During this process, a total of 193 videos were excluded: 36 due to sponsorship, 97 due to unspecified usage duration, and 60 for being demos, reviews, interviews, or first-time experiences. After exclusions, \textbf{112} videos remained for analysis. These videos had a combined duration of 26 hours and 53 minutes, ranging from 2 minutes to 1 hour and 4 minutes.

\subsubsection{Metadata Acquisition}

Metadata for these videos, including the upload date, video duration, view count, number of comments, uploading account (channel name), channel subscriber count, and country of origin, were extracted using a Python script with the YouTube Data API (see \autoref{appendix:scripts}).

\subsection{Data Analysis Strategy}

We used Qualitative Content Analysis~\cite{hsieh2005three}, one of the most frequently used methods to analyse data in digital ethnography~\cite{blythe2009critical, nielsen2023using}. In building the initial coding framework, the first coder selected a representative subset of videos, varying in duration of use, devices, and use cases. From this subset and relevant literature, key topics were identified to form the foundation of the coding framework. The framework comprises the following dimensions for classifying and analysing YouTube videos: \textit{Type of Device, Type of Channel, Duration of Use, Video Content Type, Use Cases Categories, Context of Use, Applications, User Experiences}. Most of these dimensions were developed inductively from the data, while others were adapted deductively from relevant literature. Specifically, the context dimension was informed by \citet{nielsen2023using} (environments), \citet{lu2022exploring} (changing contexts), and \citet{davari2022validating} (social contexts). 

Two coders applied the initial coding framework to a sample constituting 10\% of the video dataset. Any confusion or disagreement was resolved through discussion, leading to revisions in the framework before its application to the entire dataset. For instance, in the \textit{Context of Use}, rather than specifying environments like home, office, shopping mall, or restaurant, we categorised them by broader characteristics, such as public or private. This change allowed us to capture the overarching contextual factors influencing use, rather than focusing on specific locations. Similarly, for \textit{User Experiences}, instead of categorising experiences by Valence (Positive, Negative, or Neutral), we focused on recording quotes that reflected strong positive sentiments. This highlights particularly noteworthy experiences rather than diluting them by including neutral or negative ones, which might be less relevant for the study's objectives. The final coding framework is available in \autoref{appendix:codingframework}.

\section{Results}

The section begins by examining the dataset, focusing on, \textbf{Device}, \textbf{User/Channel}, and \textbf{Video} characteristics (Sections \ref{sec:device}, \ref{sec:user-channel}, \ref{sec:video}). This foundational analysis sets the context for understanding the rest of the results. Section \ref{sec:purpose} outlines the \textbf{Purpose of Use}, highlighting popular use cases, enablers, and applications of AR devices. Section \ref{sec:contexts} delves into the \textbf{Context of Use} in which these devices are integrated into daily life. Lastly, Section \ref{sec:compelling} \textbf{Compelling Experiences} features special instances that stood out for wearable AR users. 



\subsection{Device Characteristics} 
\label{sec:device}

\begin{table*}
  \caption{Summary of devices included in the analysis.}
  \label{device_summary}
  \small
  \renewcommand{\arraystretch}{1.5} 
  \resizebox{\linewidth}{!}{
  \begin{tabular}{lp{2.5cm}p{2.5cm}p{2.5cm}p{2.5cm}p{2.5cm}}
    \toprule
    & \textbf{\emph{Meta Ray-Ban}} & \textbf{\emph{XReal Air/Air 2}} & \textbf{\emph{Viture One/Pro}} & \textbf{\emph{Meta Quest 3}} & \textbf{\emph{Apple Vision Pro}} \\
    \midrule
    \textbf{Image} &
    \begin{minipage}[b]{0.18\columnwidth}
        \centering
        \raisebox{-.5\height}{\includegraphics[width=\linewidth]{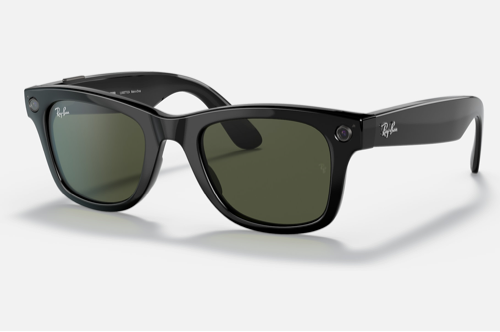}}
    \end{minipage} &
    \begin{minipage}[b]{0.18\columnwidth}
        \centering
        \raisebox{-.5\height}{\includegraphics[width=\linewidth]{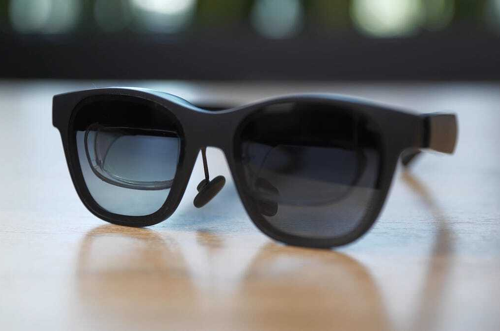}}
    \end{minipage} &
    \begin{minipage}[b]{0.18\columnwidth}
        \centering
        \raisebox{-.5\height}{\includegraphics[width=\linewidth]{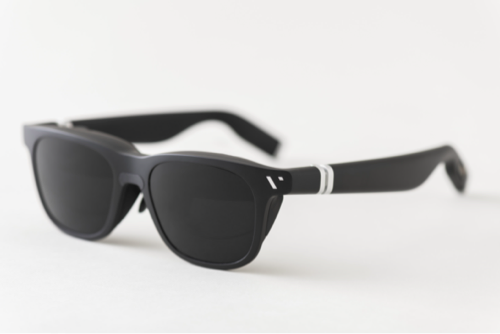}}
    \end{minipage} &
    \begin{minipage}[b]{0.18\columnwidth}
        \centering
        \raisebox{-.5\height}{\includegraphics[width=\linewidth]{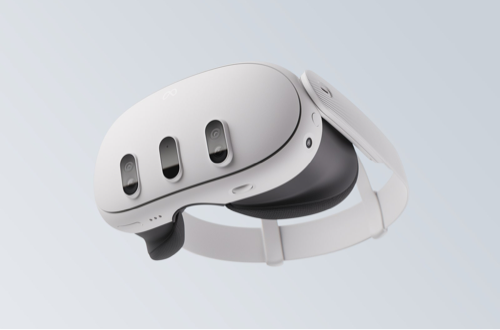}}
    \end{minipage} &
    \begin{minipage}[b]{0.18\columnwidth}
        \centering
        \raisebox{-.5\height}{\includegraphics[width=\linewidth]{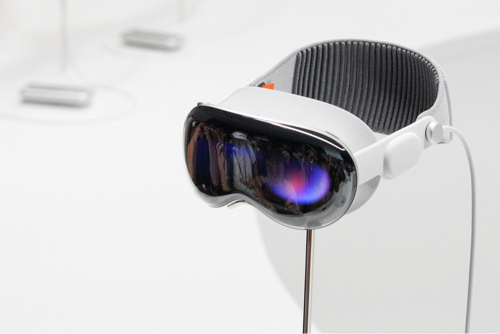}}
    \end{minipage} \\
    
    \addlinespace[0.8mm]
    \textbf{Visual Augmentation} & \cellcolor{blue!15}Audio only & \multicolumn{2}{c} {\cellcolor{teal!15}Basic augmented display} & \multicolumn{2}{c}{\cellcolor{purple!15}MR with immersive capabilities} \\

    \textbf{Device Type} & Glasses & Glasses & Glasses & Headset & Headset \\

    \textbf{Release Date} & Oct 2023 & Sep 2022/Mar 2024 & Dec 2022/May 2024 & Oct 2023 & Feb 2024 \\
    
    \textbf{Pricing (USD)} & \$299 & \$339/\$379 & \$479/\$599 & \$499 & \$3,499 \\
    
    \textbf{Company/Manufacturer} & Meta & XReal & Viture & Meta & Apple \\
    
    \textbf{Number of Videos} & 21 (18.8\%) & 6 (5.4\%) & 2 (1.8\%) & 19 (16.9\%) & 64 (57.1\%) \\
    \bottomrule
  \end{tabular}
  }
\end{table*}

AR headsets and smart glasses can be viewed as existing on a spectrum in terms of the experiences they offer to consumers (see \autoref{device_summary}).

\textit{Audio-only}: 18.8\% of videos, including the Meta Ray-Ban smart glasses. These devices primarily focus on camera and audio capabilities, such as capturing first-person videos, taking photos, and providing access to voice assistants. They do not offer any visual augmentation. 

\textit{Basic augmented display}: 7.2\% of videos, including XReal and Viture glasses. These offer more advanced visual features by integrating with existing devices like smartphones, tablets, or gaming consoles. The glasses function as an additional screen, providing an augmented display for gaming, media consumption, or productivity. While it adds a layer of visual enhancement, it is still somewhat limited compared to more advanced AR systems. 

\textit{MR with immersive capabilities}: 74\% of videos feature devices like the Apple Vision Pro (AVP) and Meta Quest 3, which use video see-through technology to blend digital content with the real world and offer fully virtual experiences. These headsets enable spatial computing, 3D object interaction, and environmental awareness.

\subsection{User/Channel Characteristics}
\label{sec:user-channel}

The content creators predominantly appeared to be male (90.2\%, 101), with female creators accounting for 9.8\%~(11). Gender was assessed based on visual appearance alone. These individuals shared their experiences through their YouTube channels, which varied in subscriber count, geographical origin, and content focus.

\textit{Subscriber count}: Following the classification by \citet{nielsen2023using}, channels were grouped based on subscriber count as either small or large. Small channels, defined as having fewer than 40,000 subscribers, accounted for 56 videos (50\%), with an average of 7,996 subscribers (SD = 11,396; Min = 4; Max = 39,300). Large channels, with more than 40,000 subscribers, also contributed 56 videos (50\%), averaging 2,582,473 subscribers (SD = 6,638,019; Min = 41,000; Max = 34,500,000). Most videos in the dataset originated from unique channels, though a few channels contributed multiple videos. For example, \textit{6 Months Later} provided two videos focusing on the AVP and Quest 3, while \textit{DeveloperAdam} explored the Quest 3 and XReal Air 2 Glasses in two videos.

\textit{Geographical origin}: The majority of the channels originated from North America, with 58\% (65) based in the United States and 8.9\% (10) in Canada. For 19.6\% (22), the country of origin was unspecified. Remaining channels were from the United Kingdom (5.4\%, 6), and individual channels from Germany, Spain, Hong Kong, Italy, the Netherlands, New Zealand, Poland, Thailand, and Uganda (0.9\% each).

\textit{Content focus}: The channels primarily focused on technology-related topics, with 52.7\% (59) dedicated to tech-focused content, such as product reviews and brand discussions. VR/AR/MR made up 9.8\% (11), while creative topics like photography, videography, and filmmaking accounted for 6.3\% (7). Miscellaneous content represented 13.4\% (15), and channels with unclear or ambiguous focus accounted for 11.6\% (13). Smaller proportions focused on gaming (3.6\%, 4) and lifestyle and wellness (2.7\%, 3).

\subsection{Video Characteristics}
\label{sec:video}

\subsubsection{Duration of Use}

\begin{figure}[htbp]
  \centering
  \includegraphics[width=1\linewidth]{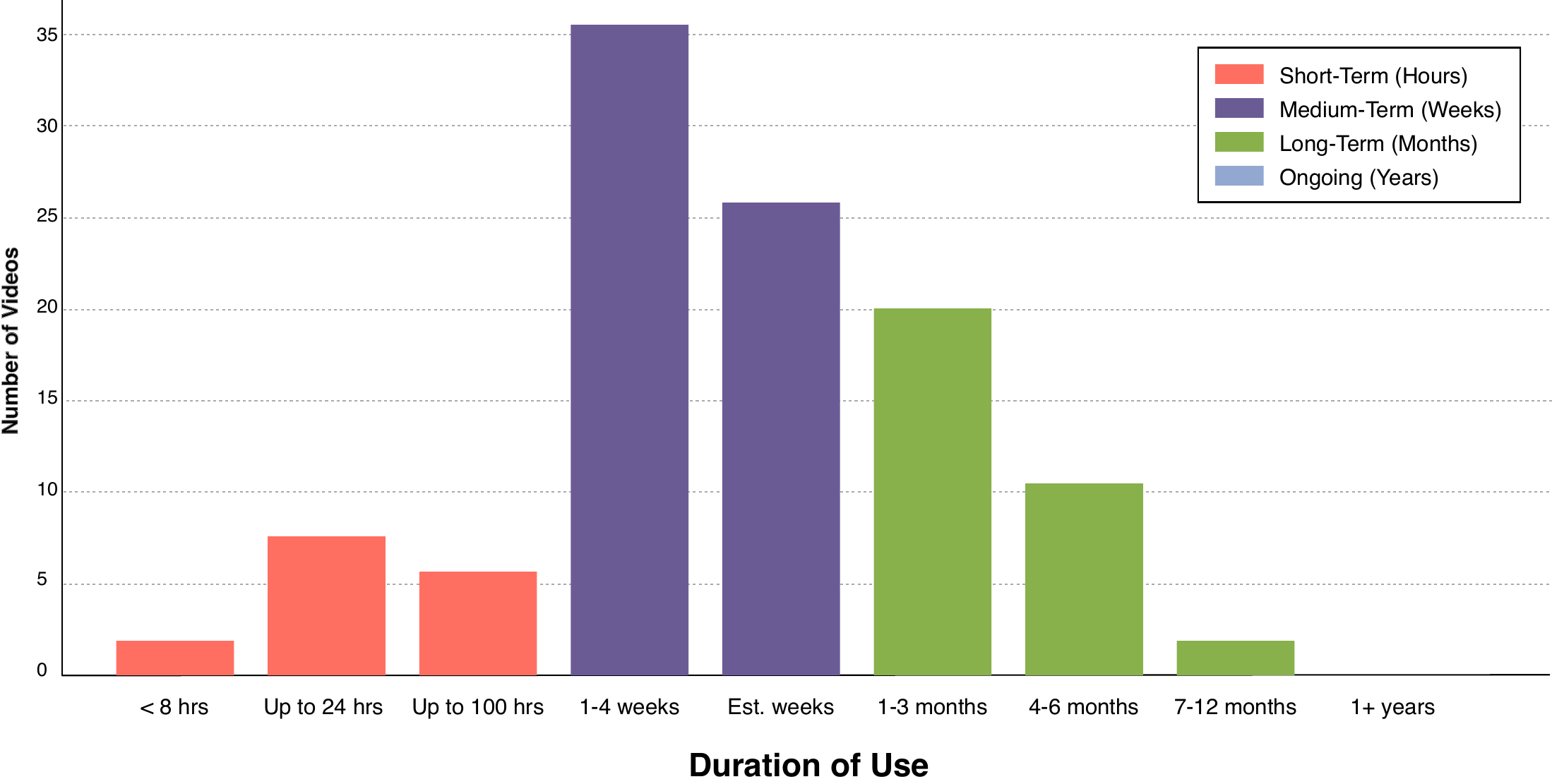}
  \caption{The chart categorises the duration of device use at the time of video creation.}
  \Description{The chart represents the distribution of videos based on the duration of device use at the time of video creation. It is divided into four categories: Short-Term (Hours), Medium-Term (Weeks), Long-Term (Months), and Ongoing (Years). The x-axis shows different time intervals, ranging from less than 8 hours to over 1 year, while the y-axis indicates the number of videos in each category, with values increasing in increments of 5 from 0 to 35. Short-term usage (less than 8 hours and up to 100 hours) dominates, with around 30 videos in this range. The number of videos decreases as the duration increases, with fewer videos depicting medium-term use (1-4 weeks and estimated weeks) and even fewer for long-term and ongoing use (up to 1+ years).}
  \label{fig:chart_duration}
\end{figure}

The videos in this study were categorised based on how long users had been using the device \textit{at the time they created the video}, except for the Short-Term Continuous Use (see \autoref{fig:chart_duration}).

\textit{Short-term continuous use (hours)}: 13.4\% of the videos, including instances where users continuously used the device for short durations, such as up to 24 or 100 hours. These videos typically involve challenges where users push the device’s limits over a short but continuous timeframe.

\textit{Medium-term use (weeks)}: 56.2\% of the videos, this category represents users who had been using the device for 1-4 weeks or provided estimates of usage spanning several weeks. 

\textit{Long-term use (months)}: 30.4\% of the videos, this category includes users who had been using the device for anywhere from 1 to 12 months. It is worth noting that most devices had been on the market for less than a year, which limits the availability of data on longer usage periods.

\renewcommand{\arraystretch}{1.2} 
\begin{table*}[ht]
    \centering
    \small
    \caption{Classification of videos based on content types.}
    \begin{tabular}{p{4.2cm}p{7cm}p{1.3cm}p{1.3cm}}
        \textbf{Content Type} & \textbf{Nature of Use} & \textbf{Authenticity} & \textbf{ Videos (\%)} \\
        \toprule
        \textbf{Using the device for a continuous period of time} & Focus on extreme use cases, where users challenge themselves to wear the device non-stop for an extended duration (e.g., a full day), even in situations where the device might not typically be used, such as during sleep. & Low-Moderate & 15 (13.4\%) \\
        \midrule
        \textbf{Using the device realistically during a day} & Depict a more balanced approach, where users integrate the device into their routine as needed. Unlike the extreme use cases, the device is used only when expected to be appropriate for specific tasks. & High & 15 (13.4\%) \\
        \midrule
        \textbf{Reflecting on the device after a certain period of use} & Provide insights after prolonged usage, highlighting technical aspects, comfort, and the user’s overall experience with the device. They focus on how the device has been integrated into daily life over time, providing a reflective analysis. & High & 82 (73.2\%) \\
        \bottomrule
    \end{tabular}
    \label{tab:content_types}
\end{table*}

\subsubsection{Video Content Type}

The videos in this study were classified based on how users documented the interaction of wearable AR devices in their daily routines (see \autoref{tab:content_types}). We also assessed the authenticity level, which indicated how closely the documented use reflected realistic, everyday interactions. While some videos with low to moderate authenticity were included, they provided valuable insights into user expectations, such as continuous access to information and interactions throughout the day. These extreme use cases demonstrated the potential for AR to become pervasive in everyday life, particularly if current technical and form factor challenges were resolved.

\subsection{Purpose of Use}
\label{sec:purpose}

This section reports on usage in key areas, including \textbf{use cases}, \textbf{enablers}, and \textbf{applications}, with a specific focus on \textbf{work-related use} in a subset of videos.

\subsubsection{Use Cases} \label{sec:use-cases}

We classified the purposes of use into four main categories and ten sub-categories, using a bottom-up approach and insights from related surveys~\cite{xraispotlight2024, krikorian2024} (see \autoref{tab:purposeful_use_combined}). The analysis showed that \textit{media consumption} (29.7\%) and \textit{gaming} (15.9\%) dominated, indicating that entertainment and leisure activities were the dominant use cases for AR in daily life. This pattern was consistent across devices, where media consumption and gaming consistently ranked as the top one or two use cases.

\textit{Work enhancement} (13.4\%) was primarily associated with fully immersive MR devices, accounting for 20.5\% of use cases on the Quest 3 and 15.3\% on the AVP. The AVP also stood out as the only device with a diverse range of productivity-related use cases, including not just work enhancement but also \textit{information retrieval}, \textit{home management}, and \textit{personal organisation}.

\textit{Media capture} was the most reported use case (40.9\%) for Meta Ray-Ban glasses, a feature rarely mentioned for other devices. \textit{Communication}, primarily through audio calls, was another prominent use case for Meta Ray-Ban (22.7\%), with users praising the open-ear speakers. AVP supported video communication via FaceTime (14.1\%), though users reported mixed experiences with its spatial persona feature. While \textit{Health and wellness} applications were less common, they were gaining attention, particularly for guided workouts and mental well-being practices.

\renewcommand{\arraystretch}{1.05}
\begin{table*}[htbp]
    \centering
    \small
    \caption{Use case categories across different devices.}
    \begin{tabular}{p{3.5cm}cccccc}
        \toprule
        \textbf{Category*} & \textbf{Total (\%)} & \textbf{Ray-Ban (\%)} & \textbf{XReal (\%)} & \textbf{Viture (\%)} & \textbf{Quest 3 (\%)} & \textbf{AVP (\%)} \\
        \midrule
        \addlinespace[1.5mm]
        \textbf{Entertainment \& Leisure} & & & & & & \\
        Media Consumption & \cellcolor{red!30}82 (29.7\%) & \cellcolor{orange!30}13 (29.5\%) & \cellcolor{orange!30}5 (38.5\%) & \cellcolor{red!30}2 (66.7\%) & \cellcolor{orange!30} 8 (20.5\%) & \cellcolor{red!30}54 (30.5\%) \\
        Gaming & \cellcolor{orange!30}44 (15.9\%) & - (-\%) & \cellcolor{red!30}6 (46.2\%) & \cellcolor{orange!30}1 (33.3\%) & \cellcolor{red!30}17 (43.6\%) & (11.3\%) \\
        \addlinespace[1.5mm]
        \textbf{Productivity \& Utility} & & & & & & \\
        Work Enhancement & \cellcolor{yellow!30}37 (13.4\%) & 1 (2.3\%) & \cellcolor{yellow!30}1 (7.7\%) & 0 (0.0\%) & \cellcolor{orange!30}8 (20.5\%) & \cellcolor{orange!30}27 (15.3\%) \\
        Information Retrieval & 24 (8.7\%) & 2 (4.5\%) & 0 (0.0\%) & 0 (0.0\%) & \cellcolor{yellow!30}2 (5.1\%) & 20 (11.3\%) \\
        Home Management & 15 (5.4\%) & 0 (0.0\%) & 0 (0.0\%) & 0 (0.0\%) & 1 (2.6\%) & 14 (7.9\%) \\
        Personal Organisation & 6 (2.2\%) & 0 (0.0\%) & 0 (0.0\%) & 0 (0.0\%) & 1 (2.6\%) & 5 (2.8\%) \\
        \addlinespace[1.5mm]
        \textbf{Social Interaction} & & & & & & \\
        Media Capture & 20 (7.2\%) & \cellcolor{red!30}18 (40.9\%) & 0 (0.0\%) & 0 (0.0\%) & 0 (0.0\%) & 2 (1.1\%) \\
        Communication & 35 (12.7\%) & \cellcolor{yellow!30}10 (22.7\%) & 0 (0.0\%) & 0 (0.0\%) & 0 (0.0\%) & \cellcolor{yellow!30}25 (14.1\%) \\
        \addlinespace[1.5mm]
        \textbf{Health \& Wellness} & & & & & & \\
        Guided Workout & 7 (2.5\%) & 0 (0.0\%) & \cellcolor{yellow!30}1 (7.7\%) & 0 (0.0\%) & 1 (2.6\%) & 5 (2.8\%) \\
        Mental Well-being & 6 (2.2\%) & 0 (0.0\%) & 0 (0.0\%) & 0 (0.0\%) & 1 (2.6\%) & 5 (2.8\%) \\
        \midrule
        \textbf{\textbf{Total}} & \textbf{276 (100.0\%)} & \textbf{44 (100.0\%)} & \textbf{13 (100.0\%)} & \textbf{3 (100.0\%)} & \textbf{39 (100.0\%)} & \textbf{177 (100.0\%)} \\
        \bottomrule
\addlinespace
\multicolumn{7}{p{14cm}}{\textsuperscript{*}Colour codes indicate the top 3 use cases for each device: red for the top 1, orange for the top 2, and yellow for the top 3.}
    \end{tabular}
    \label{tab:purposeful_use_combined}
\end{table*}

\subsubsection{Enablers} \label{sec:enablers}

The analysis revealed several key enabling features such as Virtual Monitor, Multiple Windows, and AI Assistants, which supported various use cases but were not standalone purposes themselves (see \autoref{tab:enablers}). The Virtual Monitor enabled users to project resizeable screens for tasks like work or multitasking, often requiring connection to a separate device (e.g., laptop, desktop). Multiple Windows facilitated the management of several virtual windows simultaneously. AI Assistants provided voice-enabled support for functions such as information retrieval or window management (e.g., AVP users could ask Siri to close or open applications).

\renewcommand{\arraystretch}{1.05}
\begin{table*}[htbp]
    \centering
    \small
    \caption{Enablers used across different devices.}
    \begin{tabular}{p{3.8cm}cccccc}
        \toprule
        \textbf{Category} & \textbf{Total (\%)} & \textbf{Ray-Ban (\%)} & \textbf{XReal (\%)} & \textbf{Viture (\%)} & \textbf{Quest 3 (\%)} & \textbf{AVP (\%)} \\
        \midrule
        \addlinespace[1.5mm]
        Virtual Monitor & 55 (51.4\%) & - (-\%) & 4 (57.1\%) & 1 (100.0\%) & 12 (63.2\%) & 38 (54.3\%) \\
        Multiple Windows & 35 (32.7\%) & - (-\%) & 3 (42.9\%) & 0 (0.0\%) & 7 (36.8\%) & 25 (35.7\%) \\
        AI Assistants & 17 (15.9\%) & 10 (100.0\%) & 0 (0.0\%) & 0 (0.0\%) & 0 (0.0\%) & 7 (10.0\%) \\
        \midrule
        \textbf{Total} & \textbf{107 (100.0\%)} & \textbf{10 (100.0\%)} & \textbf{7 (100.0\%)} & \textbf{1 (100.0\%)} & \textbf{19 (100.0\%)} & \textbf{70 (100.0\%)} \\
        \bottomrule
    \end{tabular}
    \label{tab:enablers}
\end{table*}

\begin{table*}[htbp]
\centering
\small
\caption{Top 20 applications mentioned.}
\begin{tabular}{clrll}
\toprule
\# & \textbf{Application} & \textbf{Mentions (\%)} & \textbf{Description} & \textbf{Mentioned Device \& Access*} \\
\midrule
1  & YouTube        & 37 (9.97\%)  & Video sharing and streaming & AVP (Browser), Quest (Native), XReal (Device) \\
2  & Safari         & 28 (7.55\%)  & Web browsing & AVP (Native) \\
3  & Environments   & 26 (7.01\%)  & Immersive experiences & AVP (Native), Quest (Native) \\
4  & FaceTime       & 16 (4.31\%)  & Video calling & AVP (Native) \\
5  & Meta AI       & 11 (2.96\%)  & AI assistant & Quest (Native), Ray-Ban (Native) \\
6  & Disney Plus    & 11 (2.96\%)  & Disney content streaming & AVP (Native) \\
7  & Siri           & 9 (2.43\%)   & Voice assistant & AVP (Native) \\
8  & Apple TV       & 9 (2.43\%)   & Video streaming & AVP (Native) \\
9  & Timer          & 7 (1.89\%)   & Built-in timer & AVP (Native) \\
10 & Messages       & 7 (1.89\%)   & Text messaging & AVP (Native), Ray-Ban (Native) \\
11 & Twitter/X      & 6 (1.62\%)   & Social media & AVP (Native) \\
12 & TikTok         & 6 (1.62\%)   & Short-form videos & AVP (Native) \\
13 & Maps           & 6 (1.62\%)   & Navigation and mapping & AVP (Native) \\
14 & Calendar       & 5 (1.35\%)   & Scheduling and management & AVP (Native) \\
15 & Zoom           & 4 (1.08\%)   & Video conferencing & AVP (Native) \\
16 & WhatsApp       & 4 (1.08\%)   & Messaging and calls & AVP (Browser), Quest (Native), Ray-Ban (Native) \\
17 & Notes          & 4 (1.08\%)   & Note-taking & AVP (Native) \\
18 & Netflix        & 4 (1.08\%)   & Video streaming & AVP (Browser), XReal (Browser) \\
19 & Fruit Ninja    & 4 (1.08\%)   & Arcade game & AVP (Native) \\
20 & Final Cut Pro  & 4 (1.08\%)   & Video editing & AVP (Device) \\
\bottomrule
\addlinespace
\multicolumn{5}{p{14cm}}{\textsuperscript{*}\textit{Native}: The application runs directly on the AR device. \textit{Browser}: The application is accessed through a web browser. \newline \textit{Device}: The application is used via a paired or mirrored device (e.g., Mac, smartphone).}
\end{tabular}
\label{tab:topapps}
\end{table*}

\subsubsection{Applications} \label{sec:applications}

The analysis documented 371 mentions of software applications across the videos, representing 138 unique applications. For the top 20 applications, see \autoref{tab:topapps}. These applications spanned several categories, including media consumption, productivity, social media and communication, immersive experiences, and hands-free interactions with voice assistants. Based on this table, several trends and patterns emerged:

\paragraph{\textbf{Overview of top applications}} The AVP emerged as the most versatile AR device on the list, supporting applications across streaming, social media, productivity, and communication. While native support for Apple's ecosystem, such as Safari and FaceTime, was expected, several third-party applications had also been optimised for the device’s AR capabilities, including TikTok, Disney Plus, and Zoom. Other third-party applications, such as YouTube and Netflix, were accessed through browsers rather than native versions. Additionally, some professional tools still depended on external setups, with applications like Final Cut Pro requiring connection to a Mac. For Meta Ray-Ban glasses, certain applications, including Meta AI and WhatsApp, were designed for voice-based interactions.

The top 20 applications listed in Table \ref{tab:topapps} were heavily skewed toward AVP due to the dataset composition. In contrast, Quest 3 was widely recognised for its immersive gaming ecosystem, but no single game was mentioned frequently enough to make the overall top 20. This contrast highlighted how application popularity varied across AR/VR platforms: AVP was primarily used for work and media, Quest dominated gaming, and XReal functioned largely as an external display rather than a standalone AR platform. The most commonly mentioned applications for XReal included Steam Deck and PlayStation, suggesting that users primarily leveraged the device for console and PC gaming, as well as productivity through second-screen functionality via Nebula and Samsung DeX.

\paragraph{\textbf{The current application ecosystem}} In many videos, the users viewed the current application ecosystem as still in its early stages, with a noticeable lack of native applications designed specifically for the device. Many applications on the visionOS App Store were repurposed from iPad apps, shown in a 3D-like environment but failing to fully exploit the immersive capabilities of AVP. One user remarked, \textit{`It reminds me of when the first iPhone came out—a bunch of weird applications developers obviously just pumped out as quickly as they could'}. Popular services such as YouTube and Spotify had yet to develop AVP-compatible versions, resulting in issues like applications crashing when used for extended periods, with one user noting, \textit{`if you’re in there for a while and trying to navigate, it tends to crash. This is happening in Safari.'} 

\paragraph{\textbf{MR games and applications}} MR, with their ability to blend virtual elements with the real world, were also frequently discussed. Users praised its immersive quality of MR games, with First Encounters being a standout example, transforming the player's environment into \textit{`a fortress invaded by cute little furballs.'} However, two users mentioned that its novelty tends to fade over time. One found ultimately preferred traditional VR games, using MR only for spatial awareness in physical games like boxing. Another user shared that after the first engaging experience with applications like a virtual model of NASA's Perseverance Rover, the excitement diminished, leaving many MR applications rarely reopened. Apart from gaming, several users envisioned AR glasses significantly enhancing everyday experiences. For navigation, one user imagined \textit{`maps that paint a picture of where you’re supposed to walk—maybe a line with arrows on the ground,'} offering intuitive guidance. Another major use case was real-time translation. A Ray-Ban user anticipated \textit{`real-time seamless translation of foreign texts while travelling,'} enabling smooth conversations, like \textit{`making a conversation with this taxi driver... and maybe complimenting the cook.'}

\subsubsection{Work-Related Use} \label{sec:work}

\sloppy While not the most prevalent, work-related applications of wearable AR devices were a significant and growing area of interest. Analysis of 112 videos revealed 13 focused on using AR devices for work tasks, with coverage distributed among AVP (8 videos), Quest 3 (3 videos), and XReal (2 videos). This indicated a notable interest in leveraging AR for productivity purposes.

Users across various AR devices appreciated the flexibility of large virtual displays for their laptops. One Quest 3 user noted: \textit{`I connected my Mac, so I can get on with some work. I currently have one virtual screen in front of me, which is gigantic, but everything is super clear.'} They also highly valued the multiple-screen functionality in AR devices, removing the constraints of screen space. One user highlighted, \textit{`I could have multiple PDFs open, research something, and look over at another document.'} Various examples illustrated how users set up their workspaces. For instance, one shared, \textit{`I have my Mac display in front of me, Final Cut open, a podcast above, and my calendar and stocks to the side,'} demonstrating how they organise and interact with multiple windows. Additionally, one user appreciated the convenience of voice commands, remarking, \textit{`I don’t normally use my voice to type or Siri, but I found myself saying, ``Hey Siri, open Messages,'' and it pops up immediately.'}

Users of MR headsets (i.e., the Quest 3 and AVP), appreciated the immersive environments that extend their workspaces beyond physical limitations. For some, shifting to scenic or creative virtual spaces, like \textit{`the Swiss Alps or a cosmic cave'}, provided a relaxing backdrop for work. The virtual environments also offered the flexibility to blend the real and virtual worlds. Users could maintain focus on their tasks while selectively interacting with their physical environment, such as seeing a keyboard or talking to colleagues. 

Nevertheless, the analysis revealed several barriers to sustained productivity while using headsets. One common issue was discomfort during prolonged use, with users stating that leaning forward while working makes the headset's weight more noticeable, as opposed to leaning back while consuming media. Some users also noted eye strain and headaches after extended use, with AVP encouraging breaks in their documentation. Additionally, low-light performance was problematic, with degraded visual quality and difficulty tracking hand movements due to limited camera functionality in dark environments. Other challenges included limited battery life, with the Quest requiring frequent use of a portable charger, especially during more resource-intensive tasks like running multiple tabs in Chrome or development tools. Users also highlighted the lack of precise input tools, such as mouse support, which hindered productivity tasks that require accuracy, and the absence of key productivity apps, like coding environments or Remote Desktop functionality. Finally, the use of virtual avatars in professional settings was occasionally perceived as unprofessional, which limited their acceptance in work environments.

\subsection{Context of Use} \label{sec:contexts}

\begin{figure*}[htbp]
  \centering
  \includegraphics[width=0.82\linewidth]{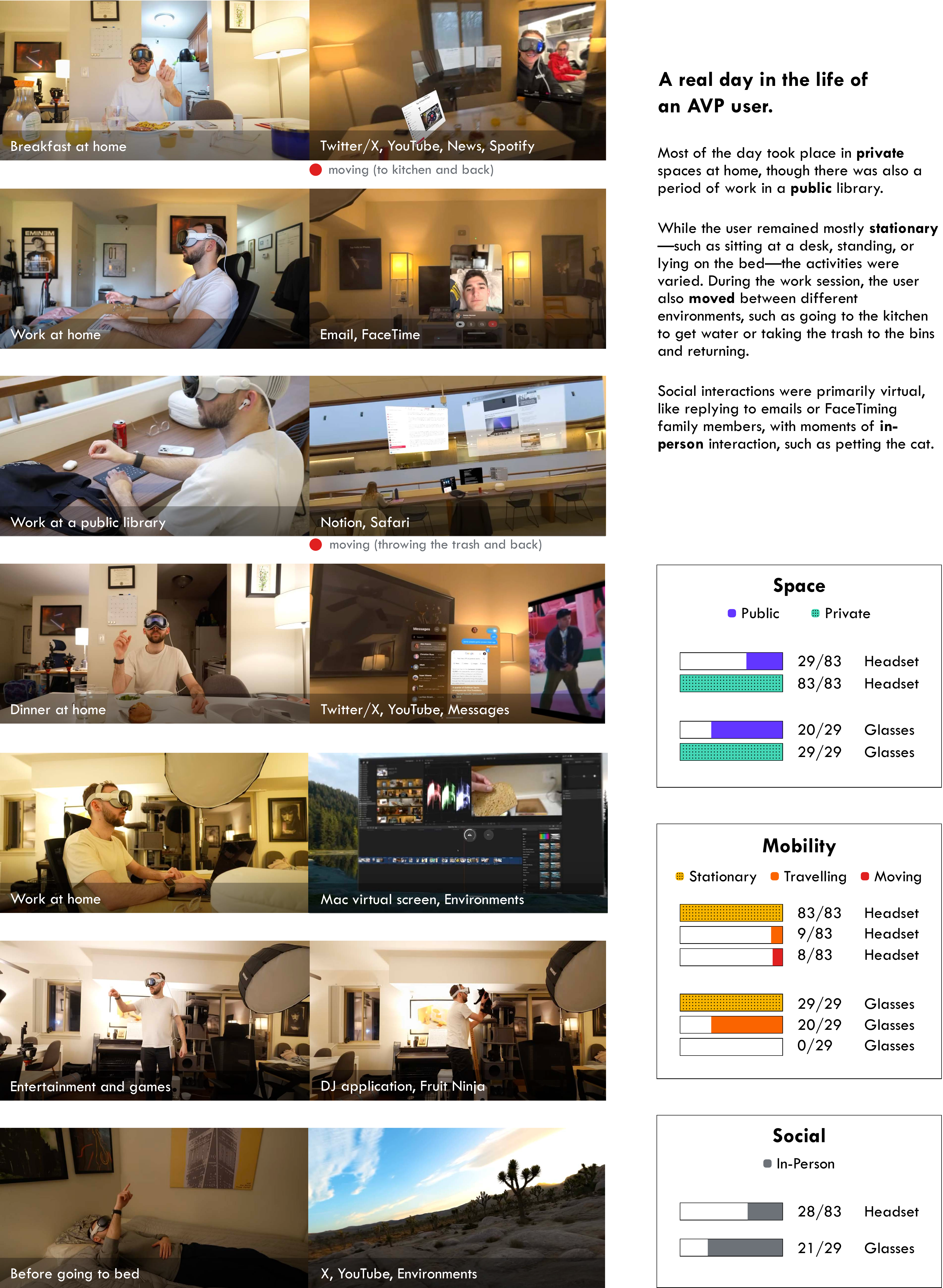}
  \caption{An example of a realistic day in the life of an AVP user, illustrating diverse applications and contexts where the headset was used. The accompanying bar chart provides an overall comparison of the AR headset and glasses usage across the entire dataset, focusing on space, mobility, and social engagement patterns.}
  \Description{This figure visually represents an example of a realistic day in the life of an AVP user, showcasing applications and contexts where the headset was used. The timeline follows a sequence of daily events, including breakfast, working from home, and working in a public library, interspersed with personal activities such as watching YouTube videos, listening to Spotify music, and using social media platforms like Twitter/X. Key transitions, such as getting water from the kitchen or taking the trash out, are also noted. Interactions with objects and pets, like petting a cat, are illustrated alongside entertainment activities such as playing Fruit Ninja and using virtual environments for productivity. Additionally, a bar chart accompanies the timeline, offering an overall comparison of AR headset and glasses usage across the dataset, focusing on three key aspects: the types of spaces users occupy, their mobility within these spaces, and their patterns of social engagement.}
  \label{fig:realistic-day}
\end{figure*}

As shown in \autoref{fig:realistic-day}, the video depicted an example of a typical day in the life of an AVP user\footnote{Apple Vision Pro - a very REAL day in the life: \url{https://www.youtube.com/watch?v=ziSnBrGdleo}}, illustrating how AR headsets are used across various contexts. The figure highlights usage instances in \textbf{space} (private vs. public), \textbf{mobility} (stationary, travelling, and moving), and \textbf{social} contexts (in-person interactions while wearing the devices). Travelling refers to moving across significant distances, such as walking, flying, or using transportation, where AR content follows the user. In contrast, moving involves shifting within smaller areas (e.g., going to the kitchen), where AR content remains spatially anchored in place and is re-engaged upon return.

Analysis of these context categories within the larger dataset, which includes both glasses and headsets, revealed distinct patterns. Headsets were overwhelmingly used in private spaces (83 of 83 instances) compared to public spaces (29 of 83). Glasses, however, had a more balanced distribution between private (29 of 29) and public (20 of 29) use. In terms of mobility, stationary activities dominated headset use (83 of 83 instances), with fewer instances of travelling (9 of 83). Glasses, by contrast, were frequently used while travelling (20 of 29 instances, primarily walking). Social interactions while wearing the devices were recorded in 28 of 83 instances for headsets and 21 of 29 instances for glasses. The following sub-sections explored these contexts in greater detail.

\paragraph{\textbf{Shifting environments temporarily}}

Users generally appreciated how AR headsets anchored content in one place, creating a stable and organised virtual workspace. This functionality allowed users to leave and return to their workspace without losing content placement. However, some users desired more flexibility, such as bringing live content with them while moving. One user found it frustrating to manually adjust the position of a sports game to keep it in view while walking.

\paragraph{\textbf{Travel mode}}

AR headsets could support travel-related contexts, such as AVP's Travel Mode, which was designed to optimise the headset for use during activities like flying or riding in a vehicle, such as a train. When enabled, the device adjusts its sensors and settings to account for the confined space and movement typically experienced during travel. However, among AVP users, some were unaware that Travel Mode requires manual activation. Once enabled, it stabilises the screen and locks it in place during head movements. Without Travel Mode, users reported that the screen could shake or glitch, particularly when the vehicle encountered bumps or turns. Additionally, the AVP was not designed for use on the go, such as while walking or driving. Three videos highlighted that the headset discourages mobile use by displaying a message alerting users: \textit{`Moving too fast—content will be hidden until you slow down.'}

\paragraph{\textbf{Social context}}

Users highlighted challenges when using AR headsets in social situations. One user described the experience as: \textit{`It’s  just you in there. It’s kind of lonely.'} Some users tried to signal that they were still aware of their surroundings by engaging in small actions, such as handing an empty soda can to a flight attendant, but they felt the device hindered their social presence. Others shared how wearing the headset led to awkward interactions, as people couldn’t make eye contact or weren’t sure if the user was available for conversation. The EyeSight feature received mixed feedback from AVP users. While some appreciated its ability to foster a sense of connection and presence in social settings, others criticised its appearance, mentioning issues such as unnatural eye placement or inconsistency depending on the user's face shape. Despite these concerns, users acknowledged its potential as a meaningful feature that could benefit from refinement in future iterations.

\paragraph{\textbf{User privacy in public}} 

Users valued the privacy features of AR devices, particularly in public spaces. For example, Viture One and Xreal Air users appreciated the ability to watch content discreetly in public places. One AVP user noted: \textit{`Now you just look like a crazy person typing away on a laptop with a totally blank screen.'} Another user valued how even small details, like typing a password, felt more secure since only they could see what was on the screen. Privacy extended beyond visual content to audio experiences. Meta Ray-Ban glasses, for instance, featured minimal sound leakage, allowing users to listen privately without disturbing others. 

\paragraph{\textbf{Bystander privacy}} 

Concerns arose regarding the visibility and effectiveness of recording indicators. For Meta Ray-Ban glasses, users noted that the small LED light designed to signal recording was difficult to notice during the day but more visible at night. Many bystanders were unaware of its purpose, as it lacked the recognisable blinking red light typically associated with recording. One user noted that covering the LED did not prevent recording, undermining its security function. Another user shared a controversial view, stating, \textit{`I really like how sneaky they are. My fiancée doesn’t love it when I take pictures with my phone, but with my glasses, she doesn’t even notice, and they feel more candid'}.

\subsection{Compelling Experiences} \label{sec:compelling}

The final part of our analysis highlighted special instances where participants were particularly impressed by features they did not expect or had not encountered before. These moments stood out not only because of their novelty but also because they represented broader themes that emerged across the videos, showing how certain applications transformed daily activities or added value in ways users had not anticipated. These instances varied significantly, driven by different levels of visual augmentation (see \autoref{special_instances}).

\begin{table}[ht]
\small
\centering
\caption{Summary of special instances.}
\begin{tabular}{p{4cm}p{3cm}}
\toprule
\textbf{Experiences} & \textbf{Visual Augmentation} \\
\midrule
(1) (Re)living Moments & \cellcolor{purple!15}Immersive VR \\ 
(2) Focusing and Escaping & \cellcolor{purple!15}Immersive VR \\ 
\addlinespace[1.3mm]
(3) Interacting Effortlessly & \cellcolor{blue!15}Audio-only \\ 
(4) Staying Present & \cellcolor{blue!15}Audio-only \\ 
\addlinespace[1.3mm]
(5) Blending Workspaces & \cellcolor{purple!15}Immersive MR \\ 
(6) Manipulating Objects Spatially & \cellcolor{purple!15}Immersive MR \\ 
(7) Augmenting Mundane Tasks & \cellcolor{purple!15}Immersive MR \\
\bottomrule
\end{tabular}
\label{special_instances}
\end{table}

\paragraph{\textbf{(1) (Re)living Moments}}
The immersive nature of videos resonated profoundly with users, blurring the line between spectator and participant. For instance, an Alicia Keys video on the AVP created a sense of presence that far exceeded traditional media experiences:

\begin{quote}
``The thing that really made me think, `Okay, this is worth it,' was the Alicia Keys video. It felt like I was in the studio with Alicia Keys and her band—it was phenomenal. If I couldn’t do anything else with the Vision Pro, I would say it’s worth it just for that!''
\end{quote}

Similarly, the use of three-dimensional photos provided a new dimension to memory recall, making participants feel as though they were reliving moments from the past:

\begin{quote}
    ``It's just breathtaking to relive that memory like this and actually see the view as I did when I was actually there.''
\end{quote}

\begin{quote}
    ``For people who may have lost a loved one or if they have a dog that is no longer here, if they wear the Vision Pro and see an old photo or video of their parents or their dog, they might start crying because it is very moving to be able to relive your memories, your visuals, in such an immersive view.''
\end{quote}

\paragraph{\textbf{(2)} Focusing and Escaping}

Users frequently emphasised the transformative experience of immersive environments, highlighting how these settings enabled them to disconnect from their physical surroundings and engage in deep work:

\begin{quote}
    ``Its biggest advantage as a productivity tool isn't the multiple windows, as you could just buy a secondary big monitor and get a similar experience. What I love about it is the focus aspect. When you put it on and turn on an environment, you are locked in—focused.''
\end{quote}

Another user spoke about the sense of escapism provided by the immersive environment, particularly in a confined space like an airplane:

\begin{quote}
    ``For the whole flight, the cramped economy cabin disappeared, transporting me to a stunning open space in Yosemite. I can’t stress enough how incredible this is.''
\end{quote}

\paragraph{\textbf{(3) Interacting Effortlessly}}

The simplicity and convenience of communication through Meta Ray-Ban’s open speaker function surprised users. Here, the focus was solely on the feeling of ease, marking a step toward a frictionless interaction with technology.

\begin{quote}
    ``Real story: I was standing in line at a store, ready to buy something, and my fiancée called. I literally answered the call by just double-tapping on the lenses. It was that quick and simple. The feeling of being able to do that was so cool!''
\end{quote}

\paragraph{\textbf{(4) Staying Present}}

Users valued staying aware of their surroundings while using the technology, whether walking, commuting, or performing everyday tasks.

\begin{quote}
    ``Do you ever go for long walks and listen to audiobooks? These glasses are perfect! You still have spatial awareness, and you can still hear the environment around you while having really good audio quality. You don’t have to worry about missing something important, like a car behind you.''
\end{quote}

Users also highlighted how the camera feature allowed them to effortlessly capture special moments, without interrupting or detracting from their present experience.

\begin{quote}
    ``They've become an extension of myself, allowing me to capture and cherish moments without sacrificing the present. For example, I was holding my daughter and enjoying the festival when an Easter Bunny flew down on a helicopter to greet us—yes, that really happened, and I managed to get it on video.''
\end{quote} 

\paragraph{\textbf{(5) Blending Workspaces}}

Multitasking across virtual spaces offered users a novel way to interact with their work environments, extending beyond traditional desktops. They viewed blending real and virtual workspaces as an evolution of their workflows, positioning AR as an extension rather than a replacement of existing technologies.

\begin{quote}
    ``Revisited the virtual space with my MacBook Pro screen shared alongside native Vision Pro apps. This is where it sort of clicked for me, and I saw a glimpse of the potential a device like the Vision Pro has as your sidekick to existing computing devices.''
\end{quote}

\paragraph{\textbf{(6) Manipulating Objects Spatially}}

When well-executed, MR enhanced spatial awareness and interaction, enabling users to manipulate objects in realistic and intuitive ways. One user cited PianoVision, showcasing how MR can make learning enjoyable while serving as an effective teaching tool.

\begin{quote}
    ``[...] coloured notes coming towards you virtually when you need to play them on the real piano.''
\end{quote}

Similarly, other users shared their enthusiasm for applications that take advantage of spatial awareness to create immersive experiences, such as 3D puzzle apps:

\begin{quote}
    ``Jigspace: You can interact with 3D models in a physical space, which is really cool. This is one of the reasons I want to use the Vision Pro more—to see more 3D models and explore how things operate.''
\end{quote}

\paragraph{\textbf{(7) Augmenting Mundane Tasks}}

One of the surprising insights from users was how MR applications were able to transform everyday, often tedious, tasks into more engaging and enjoyable experiences. One user shared their experience of using the Reminders application to manage their grocery list:

\begin{quote}
    ``I carried the Reminders application between my fingers like a grocery list. [...] It was interesting navigating the store, grabbing all my items—it felt like a grocery simulator game. Grocery shopping, which is a very mundane, boring task, became fun, exciting, and a lot more engaging.''
\end{quote}

\section{Discussion}

This section discusses the key findings of our study, contextualising them within existing literature and identifying their implications for advancing everyday AR.

\subsection{Dominant Use Cases: Entertainment and Productivity}

Our findings revealed that media consumption, gaming, and work enhancement emerged as primary use cases for wearable AR devices (Section \ref{sec:use-cases}). This trend is consistent with the results from two surveys on AVP usage. The first survey~\cite{krikorian2024}, conducted online across communities on Reddit, Facebook, LinkedIn, Twitter/X, and Threads, gathered insights from 163 AVP owners, most of whom had been using the device for 3 to 6 months. The survey revealed that 88.34\% of respondents used the device for entertainment purposes, primarily watching movies and TV shows, while 73.01\% used it for work or productivity, and 27.61\% for gaming. Similarly, a second survey~\cite{xraispotlight2024}, distributed on Twitter/X and LinkedIn, collected responses from 108 AVP users, primarily creators, developers, and founders. In this survey, 70\% of users indicated that they primarily used the device for media consumption, followed by 19\% for productivity. 

The popularity of entertainment use cases can be attributed to the immersive nature of AR technology, which enhances media consumption and gaming experiences by offering a more engaging, interactive environment compared to traditional devices. These activities require minimal setup and offer immediate benefits, making them more appealing to a wider audience. In contrast, work enhancement was less prominent due to current limitations in AR hardware, such as comfort, battery life, and integration with professional software (Section \ref{sec:work}). While these challenges make prolonged work use less practical for current users, it remains a highly desirable use cases with a lot of unrealised potential.

Resizeable virtual monitors and multiple screens are key enabling features in AR devices, enhancing work-related productivity (Section \ref{sec:enablers}). While early research~\cite{grubert2018office, pavanatto2021we} highlighted technical limitations in virtual monitors, such as usability and performance gaps compared to physical monitors, AR glasses and headsets in the current consumer market offer improved resolution, refresh rates, and overall visual quality. Complementing these ongoing hardware advancements, research—such as user studies on layout preferences for virtual workspaces during shared transit~\cite{medeiros2022shielding, ng2021airplane}, automatic layout optimisation~\cite{niyazov2023user}, and hybrid document interfaces~\cite{li2019holodoc}—could further enhance the usability and effectiveness of these enabling features once implemented. Productivity work also benefits from the immersive environments enabled by MR headsets, identified as some of the most compelling experiences in our analysis. Research by \citet{ruvimova2020transport} demonstrated that virtual environments (VE) can mitigate distractions in open office settings and enhance the ability to achieve a state of flow. While VE is traditionally associated with VR, video see-through MR headsets uniquely allow users to transition flexibly between levels of immersion tailored to their activities.

Although these enabling features support basic productivity needs, they reflect just the beginning of AR’s potential in general office work. Expanding beyond these use cases, features such as immersive data visualisation~\cite{reipschlager2021personal, lee2022designspace} and advanced collaboration tools~\cite{fender2022causality, sereno2022collaborative} align with trends in remote and hybrid work. The ability to blend physical and digital environments dynamically allows AR devices to transform workflows like meetings, brainstorming, and data analysis. These innovations, though largely in the research stage, could complement or even surpass traditional productivity paradigms as they mature. Moreover, AR devices, ranging from low- to high-end systems with varying levels of augmentation and form factors, offer tailored approaches to productivity. Audio-only glasses enhance work with hands-free communication and information access, lightweight glasses provide portable displays for basic tasks, and advanced headsets enable complex workflows. As such, productivity applications should leverage the unique strengths and account for the limitations of specific AR devices.

\begin{snugshade*} \textbf{Implications}: AR devices are establishing themselves primarily as entertainment tools. To expand their role in productivity, hardware improvements are essential to ensure comfort and enable prolonged use. Furthermore, realising AR's full potential for productivity requires moving beyond basic enabling features to create applications that fundamentally transform workflows and enhance user engagement. \end{snugshade*}

\subsection{Continuity in Digital Experience}

The analysis revealed a notable trend in how people are currently using AR headsets and smart glasses: they tended to rely on these devices for familiar applications like Netflix, YouTube, web browsing, and productivity tools (Section \ref{sec:applications}). This expectation indicated that users were primarily seeking continuity in their digital experience, leveraging AR technology to access the same functionalities they found on their smartphones, tablets, or computers.

The application ecosystem for AR devices reinforced this continuity by supporting familiar services through iPad-compatible applications or browsers. However, the current lack of native AR applications means these devices have yet to differentiate themselves by offering unique, compelling experiences that fully leverage their immersive and interactive capabilities. As one user observed, \textit{`It's kind of like being in the most state-of-the-art amusement park only to find out that there aren't a lot of rides.'} While establishing familiarity and consistency~\cite{dunser2007applying, nielsen10usability} by integrating elements from existing applications is beneficial for easing adoption, if AR headsets and smart glasses merely replicate the functionality of current devices, they may struggle to establish their value in users' daily lives beyond serving as novel alternatives.

At the same time, this expectation for continuity highlights a critical opportunity for innovation. Developers and designers have the chance to rethink how familiar applications can be re-imagined to fully exploit the strengths of wearable AR. For example, instead of simply adapting Zoom for the AVP, Zoom has created a novel `spatial experience' specifically designed for AR~\cite{zoom2024}. This new approach allows users to be represented by realistic spatial avatars, enabling others to see their facial expressions and hand gestures. Similarly, TikTok~\cite{tiktok2024} is creating an AR-specific experience, aiming to offer an immersive way for users to interact with their video feed. Innovations also extend to utility applications; for example, the Day Ahead app~\cite{dayahead2024} presents a creative method for visualising calendar events using a transparent tube filled with coloured liquid capsules, each representing different events. This unique visualisation highlights the potential for AR to transform daily tasks in engaging and practical ways.

Moving beyond continuity, the analysis revealed untapped potential for AR in other areas, such as purpose of uses with low percentage of usage (see \autoref{tab:purposeful_use_combined}). Within \textit{Productivity and Utility}, this includes general information retrieval, such as finding a car in a parking lot, ranking food items by salt content, selecting apples based on size and ripeness, and locating specific CDs in a store, as envisioned by Glassner~\cite{glassner2003everyday}. Additionally, AR has the potential to enhance personal and home management tasks, transform everyday routine activities. For instance, Dyson uses AR to show where users have vacuumed in real time~\cite{dyson2024}. In the domain of \textit{Health and Wellness}, current applications are primarily limited to guided workouts and mental wellbeing exercises, with the former often involving 2D video tutorials. However, there is significant potential to leverage spatial 3D environments and multiple viewpoints~\cite{jo2023flowar} to create more immersive and engaging experiences. By integrating sensor tracking technology and real-time contextual intelligence, future AR systems can monitor diverse metrics, including vital signs, movement patterns, and brain activity, providing personalised feedback and adapting dynamically to users' needs. For instance, tracked brain activity in learning centres may be used to adjust lesson delivery, assisting neurodivergent individuals in enhancing focus and comprehension~\cite{uspto20240090818}. These capabilities pave the way for broader health applications, bridging the gap between clinical settings and everyday health management.

\begin{snugshade*} \textbf{Implications}: 
Balancing familiarity for ease of adoption with reimagining applications to maximise AR’s strengths, such as spatial interaction and immersive design, is crucial. Untapped areas like information retrieval, home management, and healthcare offer exciting opportunities for exploration. \end{snugshade*}

\subsection{Augmenting Beyond Sight}

Compelling experiences with AR technology were identified across different levels of visual augmentation (Section \ref{sec:compelling}). While AR is often conflated with visual overlays, Milgram’s Reality-Virtuality Continuum~\cite{milgram1994taxonomy} and Azuma’s foundational definition~\cite{azuma2017making}—which emphasise combining real and virtual worlds, interactivity in real time, and 3D registration—suggest that augmentation need not be exclusively visual. Audio-first smart glasses, such as Meta Ray-Ban, straddle the line between a media device and basic audio augmentation. However, their hands-free media capture and audio interactions illustrate how minimalist augmentation can enhance everyday tasks.

Audio-first devices excel in allowing users to access information while remaining engaged with their surroundings. In our analysis, users particularly valued the utility of open-ear speakers for tasks such as listening to music and taking phone calls while staying aware of their environment. These strengths—intuitive access to information and minimal disruption—can inform broader AR design. By studying how users interact with audio-first devices, researchers and designers can extract valuable insights for future AR applications. For instance, voice-based interactions during commutes or multitasking could guide the placement and timing of visual overlays to ensure they do not disrupt primary tasks. Grounding AR innovation in real-world habits will help ensure that even visually complex systems enhance rather than fracture user engagement.

Beyond these benefits, spatial audio extends the role of auditory augmentation by adding depth and directionality to sound. In devices like the AVP, spatial audio allows users to hear sound originating from the placement of virtual objects. However, in our analysis, users did not yet find this feature particularly compelling, as its perceived impact on everyday AR use remains limited. Despite this, there is significant potential for further developments in audio technology~\cite{facebook_reality_labs_2020} within AR. Future innovations could build on existing research and advancements to enhance auditory experiences, including features like real-time language translation~\cite{rasheed2024critical}, spatial audio navigation~\cite{ruminski2015experimental}, audio guides~\cite{kwok2019gaze}, and accessibility support~\cite{li2024beyond}. These enhancements could greatly expand AR’s utility and integration into daily life, offering more inclusive and versatile applications.

\begin{snugshade*} \textbf{Implications}: Audio-only devices and spatial audio show AR’s potential to enhance everyday tasks intuitively and seamlessly. Advancing auditory experiences and leveraging user insights can broaden AR’s inclusivity and utility. \end{snugshade*}

\subsection{Towards a Safe, Ethical and Socially-Aware Integration}

\subsubsection{Ensuring Safety in Everyday Use}

The expectation that AR glasses can be seamlessly worn and used in various contexts—such as while driving, running, or engaging in other activities—highlights the importance of integrating robust safety features into these devices. For example, the AVP currently provides a warning, such as `Moving too fast,’ and deactivates digital visuals when detecting excessive movement. This proactive approach underscores the need to protect users in potentially hazardous situations. As AR devices become more pervasive, particularly those with video-see-through capabilities, addressing potential risks associated with prolonged use becomes increasingly important. There is a valuable opportunity to incorporate features similar to the Night Light mode found in other Apple products, which reduces blue light exposure in the evening. Such enhancements could significantly improve user comfort and well-being. Overall, these safety measures reflect a broader responsibility in AR design: ensuring that the technology not only enhances but also protects users' interactions with the real world.

\begin{snugshade*} \textbf{Implications}: AR devices must integrate robust safety features to ensure user well-being and safe operation across diverse contexts. \end{snugshade*}

\subsubsection{Addressing Ethical Concerns: Bystander Privacy}

\sloppy Camera recording is just one aspect of the broader concerns around bystander privacy in the context of AR glasses, which can capture or infer various forms of data, such as internal states, physiological information, and altered perceptions of reality~\cite{o2023privacy}. For example, a recent project by Harvard University students demonstrated how easily the camera in smart glasses could be modified to perform facial recognition and access personal information about individuals in public, highlighting the potential for misuse~\cite{verge2024doxxing} and the fact that many people may be unaware of the extent of data collection possible with such devices~\cite{o2023privacy}. As a result, the development of privacy-enhancing technologies (PETs) is essential to protect bystander privacy and prevent the rise of a `surveillance society'~\cite{eff_augmented_privacy_2020}. In our analysis, users of Meta Ray-Ban glasses reported that the white LED recording indicator was too subtle in daylight, and its white colour did not align with the more common expectation of a red light to signal recording. Research by \citet{koelle2018led} also found that LED indicators, though commonly used to signal camera recording, often lack visibility, clarity, security, and trustworthiness. In comparison, alternatives like mechanical shutters that physically disable the camera in privacy-sensitive situations~\cite{privaceye2019} were ranked higher in terms of trust. These findings highlight the need for further research and innovation to improve PETs. An additional insight from an online survey by~\cite{o2023privacy} suggested that PETs should consider the specific type of AR activity and the user’s relationship to bystanders, enabling selective awareness and consent. This kind of adaptive consent could be particularly useful, for example, for users wanting to take candid photos of loved ones (as indicated in Section \ref{sec:contexts}).

\begin{snugshade*} \textbf{Implications}: Privacy indicators, such as visible recording signals, must be designed for clarity and trustworthiness. Future work should explore adaptive privacy technologies to protect bystanders and foster ethical AR adoption.\end{snugshade*}

\subsubsection{Navigating Privacy and Social Interaction}

The findings highlighted the advantage of AR devices in helping users maintain focus and privacy. However, they also revealed concerns about feelings of disconnection. At this stage, the use of wearable AR, particularly headsets, remains mostly individual-centric. The technology is still evolving and has yet to fully incorporate research on shared and collaborative experiences~\cite{schroder2023collaborating, fender2022causality}, both among AR users and with non-AR users. As the analysis showed, people around an AR user often could not tell whether the user was paying attention or available for interacting with those around them. This issue is even more pronounced with devices resembling regular glasses, where it’s unclear whether the user is engaged in AR or available for interaction. This lack of transparency underscores the need for further development to improve social cues and clarity in AR use.

Though evidence was limited in our analysis, wider adoption of wearable AR could shift social norms and etiquette. The expectation of constant availability may pressure users to stay engaged with their AR devices, even in social settings where this might be seen as inappropriate. This could lead to reduced face-to-face interactions as users become more immersed in virtual experiences. These personal-level effects are accompanied by societal concerns, such as information asymmetries that may worsen existing equity issues~\cite{regenbrecht2024see}. To create meaningful experiences, future AR developments must address these challenges, promoting connection and shared experiences rather than reinforcing isolation.

\begin{snugshade*} \textbf{Implications}: AR devices should enhance social signalling and minimise isolation by improving transparency around user availability and engagement levels. Additionally, it is important to ensure that AR use supports, rather than disrupts, social cohesion. \end{snugshade*}

\section{Study Limitations}

The participants in this study were self-selected, as they chose to publicly share their experiences on YouTube. This self-selection may have introduced bias, with individuals who were particularly enthusiastic or critical more likely to create content, while those with neutral or mixed experiences may have been underrepresented. 

In the dataset, AR devices from the middle group `basic augmented displays' like XReal and Viture, accounted for only 7.2\%, largely due to the exclusion of many videos based on sponsorship criteria. We acknowledge that the findings may lack insights from this group. By contrast, the AVP’s prevalence in our dataset (57.1\%) likely stemmed from its high-profile launch and the significant attention it received from both media and technology enthusiasts. While this may have skewed the analysis towards the AVP, the device stood out as the closest among current technologies to enabling consumers to envision the possibilities of spatial computing. At the time of data collection (mid-July 2024), the AVP was not widely released globally~\cite{apple2024visionpro}, resulting in most channels being North America-based. 

Finally, this study reflected the early stage of wearable AR, where aspects such as business models, advertisement integration, and risks like surveillance or manipulation were still evolving, nowhere near the speculative vision portrayed in Keiichi Matsuda's Hyper-Reality~\cite{matsuda2016hyperreality}. While the findings focused on short- to medium-term usage patterns and early adopter perspectives, they may not have fully captured long-term or broader societal impacts. Based on these insights, future research could investigate longitudinal use, diverse demographics, and evolving ethical and societal challenges, including implications for public norms, privacy, and equity.


\section{Conclusion}

The recent progress in wearable AR technology is notable, with a diverse range of devices now reaching consumers, though they are still primarily adopted by early enthusiasts and technology pioneers. Understanding how these devices are integrated into daily life is essential for expanding AR's adoption and ensuring it delivers meaningful value across various everyday contexts. As Mark Weiser once said, \textit{`The most profound technologies are those that disappear. They weave themselves into the fabric of everyday life until they are indistinguishable from it.'} Through an analysis of 112 YouTube videos from these users, we observed that wearable AR devices were largely perceived and used as entertainment tools, while productivity also stood out as a key area of interest. Moving forward, more work is needed to ensure continuity in the digital experience and to expand AR applications into areas where its advancements can add greater value to daily life. The focus should be on creating technology that, among other goals, fosters safe, ethical, and socially-aware integration.

\begin{acks}
We thank our colleagues Natalia Gulbransen-Diaz, Xinyan Yu, and Rodrigo Hernández Ramírez for their valuable feedback in shaping this paper, with special appreciation to Natalia for her belief in this work—it truly meant a lot. We are also grateful to Noah Herman for kindly granting us permission to use screenshots from his video. Finally, we extend our thanks to the anonymous reviewers for their insightful comments.
\end{acks}

\bibliographystyle{ACM-Reference-Format}
\bibliography{references}

\appendix 

\section{Devices used as search keywords}
\label{appendix:devices}

\renewcommand{\arraystretch}{1} 
\begin{table}[htbp]
  \small
  \caption{Consumer-oriented AR headsets and glasses released between June 2022 and June 2024.}
  \label{tab:devices}
  \begin{tabular}{lp{2.7cm}p{1.6cm}p{1.6cm}}   
    \toprule
    \textbf{Company} & \textbf{Device} & \textbf{Release Date} & \textbf{Retail Price} \\
    \midrule
    Amazon & Amazon Echo Frames 3 & Dec 2023 & \$269 \\
    Apple & Apple Vision Pro & Feb 2024 & \$3,499 \\
    Meta & Quest 3 & Oct 2023 & \$499 \\
    Meta & Ray-Ban Stories 2 & Oct 2023 & \$299 \\
    TCL & RayNeo X2 & Oct 2023 & \$399 \\
    TCL & RayNeo Air 2 & Nov 2023 & \$399 \\
    \midrule
    Chamelo & Music Shield & 2023 & \$250 \\
    Everysight & Maverick & 2023 & \$600 \\
    Inmo & Air 2 & Apr 2023 & \$389 \\
    Lucyd & Glasses & Apr 2023 & \$200 \\
    Rokid & Max & May 2023 & \$439 \\
    Viture & One & Dec 2022 & \$479 \\
    Viture & One Lite & Jan 2024 & \$399 \\
    Viture & Pro & May 2024 & \$599 \\
    Vuzix & Ultralite & 2023 & \$350 \\
    XReal & Air & Sep 2022 & \$339 \\
    XReal & Air 2 & Mar 2024 & \$379 \\
    XReal & Air 2 Pro & Nov 2023 & \$459 \\
    Nubia & Neovision Glass & Jun 2023 & \$500 \\
    \bottomrule
    \addlinespace
    \multicolumn{4}{p{8cm}}{{\textsuperscript{*}Information was retrieved through Google and cross-verified using Perplexity AI for accuracy.}}
  \end{tabular}
\end{table}

\section{Scripts}
\label{appendix:scripts}

The scripts used to extract metadata and transcripts from each video have been published on the Open Science Framework (OSF): \url{https://osf.io/dtw34/?view_only=96cae98de12b4e26a2f3baa7a3498344}.

The development and refinement of the scripts were supported by OpenAI's ChatGPT, which provided guidance on improving code clarity and functionality.

Both scripts require a CSV file named video\_list.csv as input. This file must contain at least two columns: Name, which specifies the title of the video, and ID, which corresponds to the YouTube video ID.

\clearpage

\section{Coding Framework}
\label{appendix:codingframework}

\renewcommand{\arraystretch}{1.5} 
\begin{table}[h]
\centering
\footnotesize
\caption{Dimension of Analysis}
\label{codingframework_table}
\begin{tabular}{p{2.2cm}p{5.6cm}}
\toprule
\textbf{Dimension} & \textbf{Details} \\ \hline

\textbf{Type of Device} & 
\textbf{Headsets}: \newline
MR Headsets (AR/VR) \newline
AR-Only Headsets \newline
\textbf{Glasses}: \newline
Audio-Only Glasses \newline
Audio and Visual Glasses \\ \hline

\textbf{Type of Channel} & 
Small Channel \newline
Large Channel \\ \hline

\textbf{Duration of Use} & 
\textbf{Short-Term Continuous Use (Hours)}: \newline
Less than 8 hours  \newline
Up to 24 hours  \newline
Up to 100 hours \newline
\textbf{Medium-Term Use (Weeks)}: \newline
1-4 weeks \newline
Estimates* for keywords: "a day", "all day", "travel" \newline
\textbf{Long-Term Use (Months)}: \newline
1-3 months \newline
4-6 months \newline
7-12 months \newline
\textbf{On-Going Use (Years)}: \newline
1+ years  \\ \hline

\textbf{Video Content Type} & 
Using the device for a continuous period of time \newline
Using the device realistically during a day \newline
Reflecting on the device after a certain period of use \\ \hline

\textbf{Use Case Categories} & 
\textbf{Productivity \& Utility}: \newline
Work Enhancement (e.g., coding, video editing, document review) \newline
Information Retrieval (e.g., checking the weather, looking up directions) \newline
Personal Organisation (e.g., managing calendars, setting reminders) \newline
Home Management (e.g., cooking using recipe guides, grocery shopping) \newline
\textbf{Entertainment \& Leisure}: \newline
Gaming (e.g., playing Fruit Ninja in AR) \newline
Media Consumption (e.g., watching videos, listening to music) \newline
\textbf{Social Interaction}: \newline
Media Capture (e.g., photo and video capture) \newline
Communication (e.g., video calls, real-time translations during conversations) \newline
\textbf{Health \& Wellness}: \newline
Fitness Tracking (e.g., tracking steps, distance) \newline
Guided Workouts (e.g., instructions for exercise routines, yoga poses, meditation) \newline
Mental Well-being (e.g., relaxation techniques, guided meditation) \newline
\textbf{Other}: \newline
For purposes that don’t fit the above categories\\ \hline

\textbf{Context of Use} & 
Private Spaces \newline
Public Spaces \newline
Stationary \newline
Flying, walking, commuting (train), driving \newline
Shifting contexts temporarily (e.g., moving between rooms with the device on) \newline
Social (in-person interaction while wearing the device) \\ \hline

\textbf{Applications} & 
[Named Applications]\\

\hline

\textbf{User Experiences} & 
[Quotes that reflected strong positive sentiment]\\

\bottomrule

\addlinespace
\multicolumn{2}{p{8cm}}{\textsuperscript{*}The duration for video titles stating `a day', `all day', or `travel', `hundreds of hours' and `long-term' was categorised as medium-term use. These terms often referred to the content of the video, such as showcasing a single day of use or travel experiences, rather than the actual duration the device was used, which could have been longer or shorter.}
\end{tabular}
\end{table}

\end{document}